\pdfoutput=1
\documentclass[12pt,a4paper]{article}
\usepackage{ifthen} % for conditional statements
\newboolean{pdflatex}
\setboolean{pdflatex}{true} % False for eps figures 

\newboolean{articletitles}
\setboolean{articletitles}{true} % False removes titles in references

\newboolean{uprightparticles}
\setboolean{uprightparticles}{false} %True for upright particle symbols

\newboolean{inbibliography}
\setboolean{inbibliography}{false} %True once you enter the bibliography

\usepackage{microtype}
\usepackage{lineno}  % for line numbering during review
\usepackage{xspace} % To avoid problems with missing or double spaces after
\usepackage{caption} %these three command get the figure and table captions automatically small

\usepackage{graphicx}  % to include figures (can also use other packages)
\usepackage{color}
\usepackage{colortbl}
\graphicspath{{./figs/}} % Make Latex search fig subdir for figures

\usepackage{amsmath} % Adds a large collection of math symbols
\usepackage{amssymb}
\usepackage{amsfonts}
\usepackage{upgreek} % Adds in support for greek letters in roman typeset

\newcommand*\patchAmsMathEnvironmentForLineno[1]{%
\expandafter\let\csname old#1\expandafter\endcsname\csname #1\endcsname
\expandafter\let\csname oldend#1\expandafter\endcsname\csname
end#1\endcsname
 \renewenvironment{#1}%
   {\linenomath\csname old#1\endcsname}%
   {\csname oldend#1\endcsname\endlinenomath}%
}
\newcommand*\patchBothAmsMathEnvironmentsForLineno[1]{%
  \patchAmsMathEnvironmentForLineno{#1}%
  \patchAmsMathEnvironmentForLineno{#1*}%
}
\AtBeginDocument{%
\patchBothAmsMathEnvironmentsForLineno{equation}%
\patchBothAmsMathEnvironmentsForLineno{align}%
\patchBothAmsMathEnvironmentsForLineno{flalign}%
\patchBothAmsMathEnvironmentsForLineno{alignat}%
\patchBothAmsMathEnvironmentsForLineno{gather}%
\patchBothAmsMathEnvironmentsForLineno{multline}%
\patchBothAmsMathEnvironmentsForLineno{eqnarray}%
}

\usepackage{hyperref}    % Hyperlinks in references
\usepackage[all]{hypcap} % Internal hyperlinks to floats.

\def\lhcb {\mbox{LHCb}\xspace}

\def\babar  {\mbox{BaBar}\xspace}
\def\belle  {\mbox{Belle}\xspace}

\def\cdf    {\mbox{CDF}\xspace}

\def\MagUp {\mbox{\em Mag\kern -0.05em Up}\xspace}

\ifthenelse{\boolean{uprightparticles}}%
{

 \def\Ppi         {\ensuremath{\uppi}\xspace}

 \def\PDelta      {\ensuremath{\Delta}\xspace}                 
 \def\PXi      {\ensuremath{\Xi}\xspace}                 
 \def\PLambda      {\ensuremath{\Lambda}\xspace}                 
 \def\PSigma      {\ensuremath{\Sigma}\xspace}                 
 \def\POmega      {\ensuremath{\Omega}\xspace}                 
 \def\PUpsilon      {\ensuremath{\Upsilon}\xspace}                 
 
 \def\PB      {\ensuremath{\mathrm{B}}\xspace}                 
                  
 \def\PD      {\ensuremath{\mathrm{D}}\xspace}

 \def\PK      {\ensuremath{\mathrm{K}}\xspace}

 \def\Pb      {\ensuremath{\mathrm{b}}\xspace}                 
 \def\Pc      {\ensuremath{\mathrm{c}}\xspace}

 \def\Pi      {\ensuremath{\mathrm{i}}\xspace}

}
{

 \def\Ppi         {\ensuremath{\pi}\xspace}

 \mathchardef\PDelta="7101
 \mathchardef\PXi="7104
 \mathchardef\PLambda="7103
 \mathchardef\PSigma="7106
 \mathchardef\POmega="710A
 \mathchardef\PUpsilon="7107
                  
 \def\PB      {\ensuremath{B}\xspace}                 
                  
 \def\PD      {\ensuremath{D}\xspace}

 \def\PK      {\ensuremath{K}\xspace}

 \def\Pb      {\ensuremath{b}\xspace}                 
 \def\Pc      {\ensuremath{c}\xspace}

 \def\Pi      {\ensuremath{i}\xspace}

}

\makeatletter
\ifcase \@ptsize \relax% 10pt
  \newcommand{\miniscule}{\@setfontsize\miniscule{4}{5}}% \tiny: 5/6
\or% 11pt
  \newcommand{\miniscule}{\@setfontsize\miniscule{5}{6}}% \tiny: 6/7
\or% 12pt
  \newcommand{\miniscule}{\@setfontsize\miniscule{5}{6}}% \tiny: 6/7
\fi
\makeatother

\DeclareRobustCommand{\optbar}[1]{\shortstack{{\miniscule (\rule[.5ex]{1.25em}{.18mm})}
  \\ [-.7ex] $#1$}}

\def\cquark    {{\ensuremath{\Pc}}\xspace}

\def\bquark    {{\ensuremath{\Pb}}\xspace}

\def\pion   {{\ensuremath{\Ppi}}\xspace}
\def\piz    {{\ensuremath{\pion^0}}\xspace}

\def\pip    {{\ensuremath{\pion^+}}\xspace}
\def\pim    {{\ensuremath{\pion^-}}\xspace}

\def\kaon    {{\ensuremath{\PK}}\xspace}
  \def\Kbar    {{\kern 0.2em\overline{\kern -0.2em \PK}{}}\xspace}

\def\KorKbar    {\kern 0.18em\optbar{\kern -0.18em K}{}\xspace}

\def\Kp      {{\ensuremath{\kaon^+}}\xspace}
\def\Km      {{\ensuremath{\kaon^-}}\xspace}

  \def\Dbar    {{\kern 0.2em\overline{\kern -0.2em \PD}{}}\xspace}
\def\D       {{\ensuremath{\PD}}\xspace}

\def\DorDbar    {\kern 0.18em\optbar{\kern -0.18em D}{}\xspace}
\def\Dz      {{\ensuremath{\D^0}}\xspace}
\def\Dzb     {{\ensuremath{\Dbar{}^0}}\xspace}

\def\Dstar   {{\ensuremath{\D^*}}\xspace}

\def\Dstarp  {{\ensuremath{\D^{*+}}}\xspace}
\def\Dstarm  {{\ensuremath{\D^{*-}}}\xspace}

\def\Bbar    {{\ensuremath{\kern 0.18em\overline{\kern -0.18em \PB}{}}}\xspace}

\def\BorBbar    {\kern 0.18em\optbar{\kern -0.18em B}{}\xspace}

  \def\Y#1S{\ensuremath{\PUpsilon{(#1S)}}\xspace}% no space before {...}!

\def\Lbar        {{\ensuremath{\kern 0.1em\overline{\kern -0.1em\PLambda}}}\xspace}
\def\LorLbar    {\kern 0.18em\optbar{\kern -0.18em \PLambda}{}\xspace}

\def\to                 {\ensuremath{\rightarrow}\xspace}

\def\order   {{\ensuremath{\mathcal{O}}}\xspace}

\def\CP                {{\ensuremath{C\!P}}\xspace}

\def\AT#1     {\ensuremath{A_{\mathrm{T}}^{#1}}\xspace}           % 2

\def\C#1      {\ensuremath{\mathcal{C}_{#1}}\xspace}                       % 9
\def\Cp#1     {\ensuremath{\mathcal{C}_{#1}^{'}}\xspace}                    % 7
\def\Ceff#1   {\ensuremath{\mathcal{C}_{#1}^{\mathrm{(eff)}}}\xspace}        % 9  
\def\Cpeff#1  {\ensuremath{\mathcal{C}_{#1}^{'\mathrm{(eff)}}}\xspace}       % 7
\def\Ope#1    {\ensuremath{\mathcal{O}_{#1}}\xspace}                       % 2
\def\Opep#1   {\ensuremath{\mathcal{O}_{#1}^{'}}\xspace}                    % 7

\def\ycp        {\ensuremath{y_{\CP}}\xspace}

\newcommand{\tev}{\ifthenelse{\boolean{inbibliography}}{\ensuremath{~T\kern -0.05em eV}\xspace}{\ensuremath{\mathrm{\,Te\kern -0.1em V}}}\xspace}
\newcommand{\gev}{\ensuremath{\mathrm{\,Ge\kern -0.1em V}}\xspace}
\newcommand{\mev}{\ensuremath{\mathrm{\,Me\kern -0.1em V}}\xspace}
\newcommand{\kev}{\ensuremath{\mathrm{\,ke\kern -0.1em V}}\xspace}
\newcommand{\ev}{\ensuremath{\mathrm{\,e\kern -0.1em V}}\xspace}
\newcommand{\gevc}{\ensuremath{{\mathrm{\,Ge\kern -0.1em V\!/}c}}\xspace}
\newcommand{\mevc}{\ensuremath{{\mathrm{\,Me\kern -0.1em V\!/}c}}\xspace}
\newcommand{\gevcc}{\ensuremath{{\mathrm{\,Ge\kern -0.1em V\!/}c^2}}\xspace}
\newcommand{\gevgevcccc}{\ensuremath{{\mathrm{\,Ge\kern -0.1em V^2\!/}c^4}}\xspace}
\newcommand{\mevcc}{\ensuremath{{\mathrm{\,Me\kern -0.1em V\!/}c^2}}\xspace}

\def\invfb   {\ensuremath{\mbox{\,fb}^{-1}}\xspace}

\newcommand{\stat}{\ensuremath{\mathrm{\,(stat)}}\xspace}
\newcommand{\syst}{\ensuremath{\mathrm{\,(syst)}}\xspace}

\def\order{{\ensuremath{\cal O}}\xspace}
\newcommand{\chisq}{\ensuremath{\chi^2}\xspace}

\def\gsim{{~\raise.15em\hbox{$>$}\kern-.85em
          \lower.35em\hbox{$\sim$}~}\xspace}
\def\lsim{{~\raise.15em\hbox{$<$}\kern-.85em
          \lower.35em\hbox{$\sim$}~}\xspace}

\newcommand{\mean}[1]{\ensuremath{\left\langle #1 \right\rangle}} % {x}
\def\tell1  {TELL1\xspace}
\def\ukl1   {UKL1\xspace}

\usepackage{cite} % Allows for ranges in citations
\usepackage{mciteplus}

\usepackage{longtable} % only for template; not usually to be used in PAPERs

\newboolean{wordcount}
\setboolean{wordcount}{false}

\newboolean{plot_for_prl}
\setboolean{plot_for_prl}{false}

\newcommand{\DACP}{\ensuremath{\Delta A_{\CP}}\xspace}
\newcommand{\KK}{\ensuremath{\Dz\to\Km\Kp}\xspace}
\newcommand{\PiPi}{\ensuremath{\Dz\to\pim\pip}\xspace}
\newcommand{\deltam}{{\ensuremath{\delta m}}\xspace}

\def\ARAW {\ensuremath{A_{\mathrm{raw}}}\xspace}
\def\AD {\ensuremath{A_{\mathrm{D}}}\xspace}
\def\AP {\ensuremath{A_{\mathrm{P}}}\xspace}

\begin{document}

%%%%%%%%%%%%%%%%%%%%%%%%%
%%%%% Title     %%%%%%%%%
%%%%%%%%%%%%%%%%%%%%%%%%%
\renewcommand{\thefootnote}{\fnsymbol{footnote}}
\setcounter{footnote}{1}

% %%%%%%% CHOOSE TITLE PAGE--------
% $Id: title-LHCb-PAPER.tex 85872 2016-01-08 15:53:15Z egersa $
% ===============================================================================
% Purpose: LHCb-PAPER journal paper title page template
% Author: 
% Created on: 2010-09-25
% ===============================================================================

%%%%%%%%%%%%%%%%%%%%%%%%%
%%%%%  TITLE PAGE  %%%%%%
%%%%%%%%%%%%%%%%%%%%%%%%%
\begin{titlepage}
\pagenumbering{roman}

% Header ---------------------------------------------------
\vspace*{-1.5cm}
\centerline{\large EUROPEAN ORGANIZATION FOR NUCLEAR RESEARCH (CERN)}
\vspace*{1.5cm}
\hspace*{-0.5cm}
\begin{tabular*}{\linewidth}{lc@{\extracolsep{\fill}}r}
\ifthenelse{\boolean{pdflatex}}% Logo format choice
{\vspace*{-2.7cm}\mbox{\!\!\!\includegraphics[width=.14\textwidth]{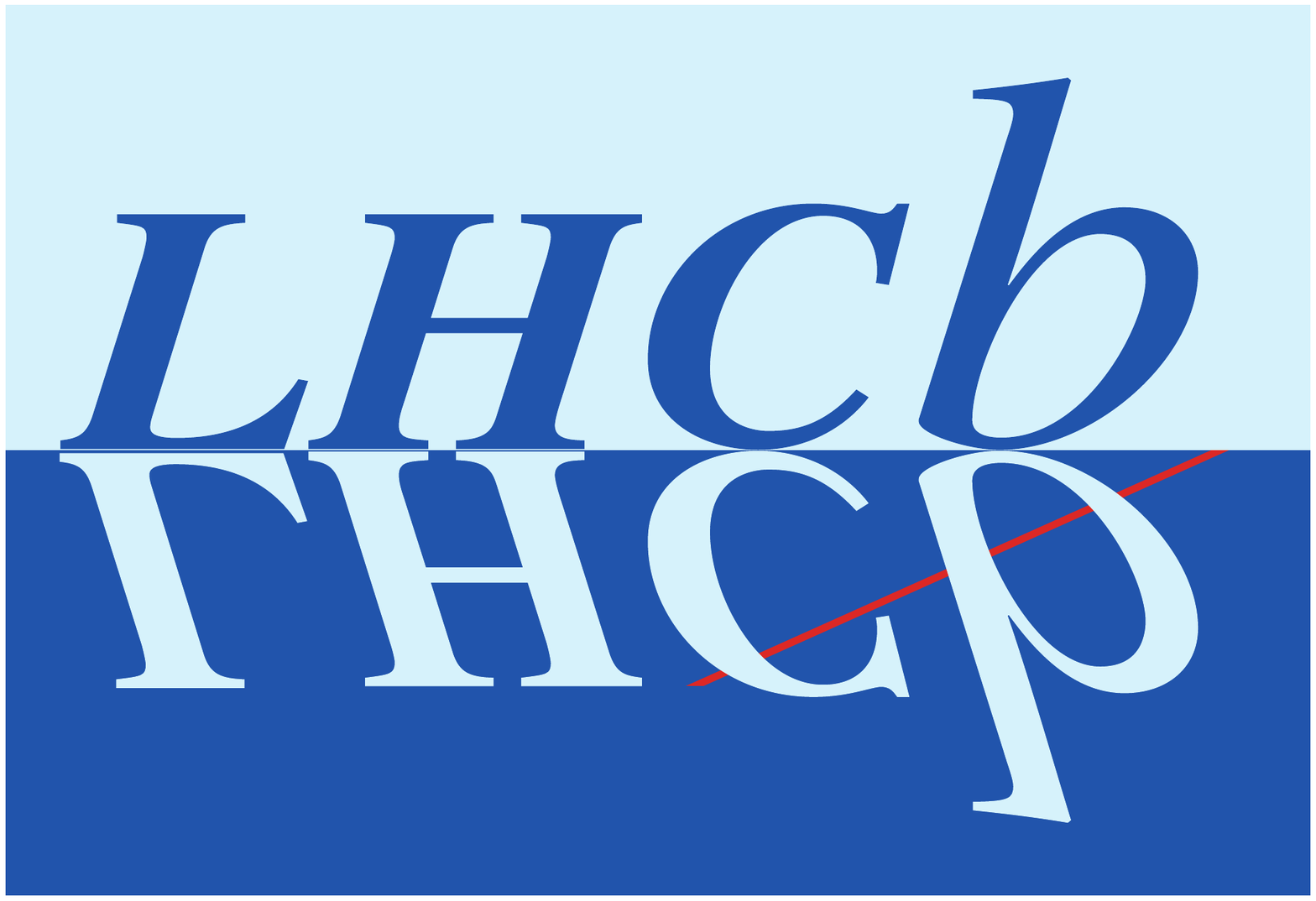}} & &}%
{\vspace*{-1.2cm}\mbox{\!\!\!\includegraphics[width=.12\textwidth]{lhcb-logo.eps}} & &}%
\\
 & & CERN-EP-2016-022 \\  % ID 
 & & LHCb-PAPER-2015-055 \\  % ID 
\end{tabular*}

\vspace*{3.0cm}

% Title --------------------------------------------------
{\bf\boldmath\huge
\begin{center}

Measurement of the difference of time-integrated \CP asymmetries in $\Dz \to K^{-} K^{+} $ and $\Dz \to \pi^{-} \pi^{+} $ decays
\end{center}
}

\vspace*{2.0cm}

\begin{center}
The LHCb collaboration\footnote{Authors are listed at the end of this letter.}
\end{center}

% Abstract -----------------------------------------------
\begin{abstract}
  \noindent
A search for  \CP violation in \KK and \PiPi decays is performed using $pp$ collision data, corresponding to an integrated luminosity of 3\invfb,  collected using the LHCb detector at centre-of-mass energies of 7 and 8\tev. The flavour of the charm meson is inferred from the charge of the pion in
$\Dstarp\to\Dz\pip$ and $\Dstarm\to\Dzb\pim$ decays.
The difference between the \CP asymmetries in \KK and \PiPi
decays, $\DACP \equiv A_{\CP}(\Km\Kp) - A_{\CP}(\pim\pip)$, is
measured to be $\left( -0.10 \pm 0.08\,\stat \pm 0.03\,\syst \right) \%$. This is the most precise measurement of a time-integrated \CP asymmetry in the charm sector from a single experiment.

\end{abstract}

\vspace*{1.0cm}

\begin{center}
  Published in Phys.~Rev.~Lett. 116, 191601 (2016)
\end{center}

\vspace{\fill}

{\footnotesize 
\centerline{\copyright~CERN on behalf of the \lhcb collaboration, licence \href{http://creativecommons.org/licenses/by/4.0/}{CC-BY-4.0}.}}
\vspace*{2mm}

\end{titlepage}

%%%%%%%%%%%%%%%%%%%%%%%%%%%%%%%%
%%%%%  EOD OF TITLE PAGE  %%%%%%
%%%%%%%%%%%%%%%%%%%%%%%%%%%%%%%%

\newpage
\setcounter{page}{2}
\mbox{~}

\cleardoublepage

% %%%%%%%%%%%%% ---------

\renewcommand{\thefootnote}{\arabic{footnote}}
\setcounter{footnote}{0}

%%%%%%%%%%%%%%%%%%%%%%%%%
%%%%% Main text %%%%%%%%%
%%%%%%%%%%%%%%%%%%%%%%%%%

\pagestyle{plain} % restore page numbers for the main text
\setcounter{page}{1}
\pagenumbering{arabic}

\noindent Violation of charge-parity (\CP) symmetry in weak decays of hadrons is described in the Standard Model (SM) by the Cabibbo-Kobayashi-Maskawa (CKM) matrix and has been observed in 
$K$- and $B$-meson systems~\cite{Christenson:1964fg,Aubert:2004qm,Chao:2004mn,LHCb-PAPER-2013-018,LHCb-PAPER-2012-001}. 
However, no \CP violation has been observed in the charm sector,
despite the experimental progress seen in charm physics in the last decade. Examples are the unambiguous observation of \Dz--\Dzb meson mixing~\cite{bib:babarmixingmoriond,Staric:2015sta,Aaltonen:2007ac,LHCb-PAPER-2012-038,LHCb-PAPER-2013-053,LHCb-PAPER-2015-057}, and measurements of \CP asymmetry observables in \D mesons decays, reaching an experimental precision of $\order(10^{-3})$~\cite{HFAG}. 
The amount of \CP violation is expected to be below the percent level~\cite{Feldmann:2012js, Bhattacharya:2012ah, Pirtskhalava:2011va,
Brod:2012ud, Cheng:2012xb, Muller:2015rna,Golden:1989qx,Li:2012cfa}, but large theoretical uncertainties due to long distance interactions prevent precise SM calculations. Charm hadrons provide a unique opportunity to search for \CP violation with particles containing only up-type quarks.

This Letter presents a measurement of the difference between the time-integrated \CP asymmetries of $D^0 \rightarrow K^-K^+$ and $D^0 \rightarrow \pi^-\pi^+$  decays, performed with $pp$ collision data corresponding to an integrated luminosity of 3\invfb collected using the LHCb detector at centre-of-mass energies of 7 and 8\tev. The inclusion of charge-conjugate decay modes is implied throughout except in the definition of asymmetries. This result is an update of the previous LHCb measurement with 0.6\invfb of data, in which a value of $\DACP = (-0.82 \pm 0.21)\%$ was obtained~\cite{LHCb-PAPER-2011-023}. 

The time-dependent \CP asymmetry, $A_{\CP}(f;\,t)$, for $D^0$ mesons decaying to a \CP eigenstate $f$ is defined as
\begin{equation}
A_{\CP}(f;\,t) \equiv \frac{\Gamma(\Dz(t) \to f)-\Gamma(\Dzb(t) \to f)}{\Gamma(\Dz(t) \to f)+\Gamma(\Dzb(t) \to f)}, \label{eq:acpf}
\end{equation}
where $\Gamma$ denotes the decay rate. For $f= K^- K^+$ and $f= \pi^- \pi^+$, $A_{\CP}(f;\,t)$ can be expressed in terms of  a direct component associated with \CP violation in the decay amplitudes, and an indirect component associated with \CP violation in the mixing or in the interference between mixing and decay. In the limit of exact symmetry under a transformation interchanging $d$ and $s$ quarks (U-spin symmetry), the direct component is expected to be equal in magnitude and opposite in sign for $K^-K^+$ and $\pi^-\pi^+$ decays~\cite{bib:grossmankagannir}. However, large U-spin breaking effects could be present~\cite{Feldmann:2012js,Gronau:2015rda,Grossman:2012ry,Brod:2012ud}.

The measured time-integrated asymmetry, $A_{\CP}(f)$, depends upon the reconstruction efficiency as a function of the decay time. It can be written as~\cite{bib:cdfpaper,Gersabeck:2011xj}
\begin{equation}
A_{\CP}(f)  \approx a_{\CP}^{\rm dir}(f) \left(1+\frac{\langle t (f)\rangle}{\tau}\,\ycp\right) + \frac{\langle t (f)\rangle}{\tau}\,a_{\CP}^{\rm ind},\label{eq:acpphysics}
\end{equation}
where $\langle t (f)\rangle$ denotes the mean decay time of \Dz\to $f$ decays in the reconstructed sample, $a_{\CP}^{\rm dir}(f)$ as the direct \CP asymmetry, $\tau$ the \Dz lifetime, $a_{\CP}^{\rm ind}$ the indirect \CP asymmetry and $\ycp$ is the deviation from unity of the ratio of the effective lifetimes of decays to flavour specific and \CP-even final states. 
To a good approximation, $a_{\CP}^{\rm ind}$ is independent of the decay mode~\cite{bib:grossmankagannir,bib:kagansokoloff}.

Neglecting terms of the order $\order(10^{-6})$, the difference in \CP asymmetries between $\Dz \to \Km\Kp$ and $\Dz \to \pim\pip$ is
\begin{eqnarray}
\DACP  & \equiv &  A_{\CP}(\Km\Kp) - A_{\CP}(\pim\pip) \nonumber \label{eq:dacpdef1} \\ 
& \approx & \Delta a_{\CP}^{\rm dir}\left(1 + \frac{\overline{\langle t \rangle}}{\tau}\,\ycp\right) + \frac{\Delta \langle t \rangle}{\tau}\,a_{\CP}^{\rm ind} ,
\label{eq:dacpdef} 
\end{eqnarray}
where $\overline{\langle t \rangle}$ is the arithmetic average of $\langle t (\Km\Kp)\rangle$ and  $\langle t (\pim\pip)\rangle$.

The most precise measurements of the time-integrated \CP asymmetries in $\Dz\to\Km\Kp$ and $\Dz\to\pim\pip$ decays to date have been performed by the \lhcb~\cite{LHCb-PAPER-2011-023,LHCb-PAPER-2014-013}, \cdf~\cite{CDF:2012qw}, \babar~\cite{bib:babarpaper2008} and \belle~\cite{bib:bellepaper2008,*Ko:2012jh} collaborations. The measurement in Ref.~\cite{LHCb-PAPER-2014-013} uses \Dz mesons produced in semileptonic $b$-hadron decays, 
where the charge of the muon is used to identify the flavour of the \Dz meson at production, while the other measurements use \Dz mesons produced in the decay of the $\Dstar(2010)^+$ meson, hereafter referred to as \Dstarp.

The raw asymmetry, $\ARAW(f)$, measured for \Dz decays to a final state $f$ is defined as
\begin{equation}
\ARAW(f) \equiv \frac {N\left(\Dstarp \to \Dz(f)\pi^{+}_{s}\right) - N\left(\Dstarm \to \Dzb(f)\pi^{-}_{s}\right)} {N\left(\Dstarp \to \Dz(f)\pi^{+}_{s}\right) + N\left(\Dstarm \to \Dzb(f)\pi^{-}_{s}\right)} , 
\label{araw}
\end{equation}
where $N$ is the number of reconstructed signal candidates of the given decay and the flavour of the $D^0$ meson is identified using the charge of the soft pion ($\pi^+_s$) in the strong decay $\Dstarp\to\Dz\pi^{+}_s$. The raw asymmetry can be written, up to $\order(10^{-6})$, as 
\begin{equation}
\ARAW(f) \approx A_{\CP}(f) + \AD(f) + \AD(\pi_s^+) + \AP(\Dstarp),
\label{def:arawstarcomponents}
\end{equation}
where $\AD(f)$ and $\AD(\pi_s^+)$ 
are the asymmetries in the reconstruction efficiencies of the \Dz final state and of the soft pion,
and $\AP(\Dstarp)$ is the production asymmetry for \Dstarp mesons, arising from 
the hadronisation of charm quarks in $pp$ collisions. The magnitudes of $\AP(\Dstarp)$~\cite{LHCb-PAPER-2012-026} and $\AD(\pi_s^+)$~\cite{LHCb-PAPER-2012-009} are both about 1\%.
Equation~\ref{def:arawstarcomponents} is only valid when reconstruction efficiencies of the final state $f$ and of the soft pion are independent. 
Since both $\Km\Kp$ and $\pim\pip$ final states are self-conjugate, $\AD(\Km\Kp)$ and $\AD(\pim\pip)$ are identically zero.
To a good approximation $\AD(\pi^{+}_s)$ and $\AP(\Dstarp)$ are independent of the final state $f$ in any given kinematic region, and thus cancel in the difference, giving
\begin{align}
\begin{split}
\DACP = \ARAW(\Km\Kp) - \ARAW(\pim\pip).
\label{DACP1}
\end{split}
\end{align}
However, to take into account an imperfect cancellation of detection and production asymmetries due to the difference in the kinematic properties of the two decay modes, 
the kinematic distributions of \Dstarp mesons decaying to the \Km\Kp final state are reweighted to match those of \Dstarp mesons decaying to the \pim\pip final state. The weights are calculated for each event using the ratios of the background-subtracted distributions of the \Dstarp momentum, transverse momentum and azimuthal angle for both final states after the final selection.

The \lhcb detector~\cite{Alves:2008zz,LHCb-DP-2014-002} is a single-arm forward spectrometer covering the \mbox{pseudorapidity} range $2<\eta <5$, designed for the study of particles containing \bquark or \cquark quarks. 
The two ring-imaging Cherenkov detectors~\cite{LHCb-DP-2012-003} provide particle identification (PID) to distinguish kaons from pions for momenta ranging from a few\gevc to about 100\gevc. The direction of the field polarity (up or down) of the LHCb dipole magnet is reversed periodically, giving data samples of comparable size for both magnet polarities.

To select \Dstarp candidates, events must satisfy hardware and software trigger requirements and a subsequent offline selection. 
The trigger consists of a hardware stage, based on high transverse momentum signatures in the calorimeter and muon systems, followed by a software stage, which applies a full event reconstruction. When the hardware trigger decision was initiated by calorimeter deposits from \Dz decay products, the event is categorised as ``triggered on signal'' (TOS). Events that are not TOS, but in which the hardware trigger decision is due to particles in the event other than the \Dstarp decay products, are also accepted; these are referred to as ``not triggered on signal'' (nTOS). 
The events associated with these trigger categories present different kinematic properties. To have cancellation of production and detection asymmetries the data are split into TOS and nTOS samples and \DACP is measured separately in each sample. 

Both the software trigger and subsequent event selection use kinematic variables and decay time to isolate the signal decays from the background. Candidate \Dz mesons  must have a decay vertex that is well separated from all primary $pp$ interaction vertices (PVs). They are combined with pion candidates to form \Dstarp candidates.  
Requirements are placed on: the track fit quality;  the $D^{*+}$ vertex fit quality, where the vertex formed by $D^0$ and $\pi_s^+$ candidates is constrained to coincide with the associated PV~\cite{Hulsbergen2005566}; the \Dz transverse momentum 
and its decay distance;
the angle between the \Dz momentum in the laboratory frame and the momentum of the kaon or the pion in the \Dz rest frame;
the smallest impact parameter chi-squared (IP $\chi^2$) of both the \Dz candidate and its decay products with respect to all PVs in the event.
The IP $\chi^2$ is defined as the difference between the $\chi^2$ of the PV reconstructed with and without the considered particle.
Cross-feed backgrounds from \D meson decays with a
kaon misidentified as a pion, and vice versa, are reduced using PID requirements.
After these selection criteria, the dominant background consists of genuine \Dz candidates paired with unrelated pions originating from the interaction vertex.

Fiducial requirements are imposed to exclude kinematic regions having a large asymmetry in the soft pion reconstruction efficiency (see Figs.~\ref{fig:fidcuts1} and \ref{fig:fidcuts2} in Ref.~\cite{supplemental}).
These regions occur because low momentum particles of one charge at large (small) angles in the horizontal plane may be deflected out of the detector acceptance (into the non-instrumented beam pipe region) whereas particles with the other charge are more likely to remain within the acceptance. About 70\% of the selected candidates are retained by these fiducial requirements.

The candidates satisfying the selection criteria are accepted for further analysis if the mass difference $\delta m \equiv  m(h^+h^-\pi^+_s) -m(h^+h^-) - m(\pi^+)$  for $h=K,\pi$ is in the range 0.2--12.0\mevcc and the invariant mass of the \Dz candidate is within two standard deviations from the central value of the mass resolution model. The standard deviation corresponds to about 8\mevcc and 10\mevcc for \KK and \PiPi decays, respectively.

The data sample includes events with multiple \Dstarp candidates.
The majority of these events contain the same reconstructed \Dz meson combined with different soft pion candidates. The fraction of events with multiple candidates in a range of \deltam corresponding to 4.0--7.5\mevcc is about 1.2\% for TOS events and 2.4\% for nTOS events; these fractions are the same for the $K^-K^+$ and $\pi^-\pi^+$ final states, and for both magnet polarities. The events with multiple candidates are retained and a systematic uncertainty is assessed. 

\begin{figure}[tb]
\begin{center}
\ifthenelse{\boolean{plot_for_prl}}
{
\includegraphics[width=0.48\textwidth]{Fig1.pdf}
}{
\includegraphics[width=0.48\textwidth]{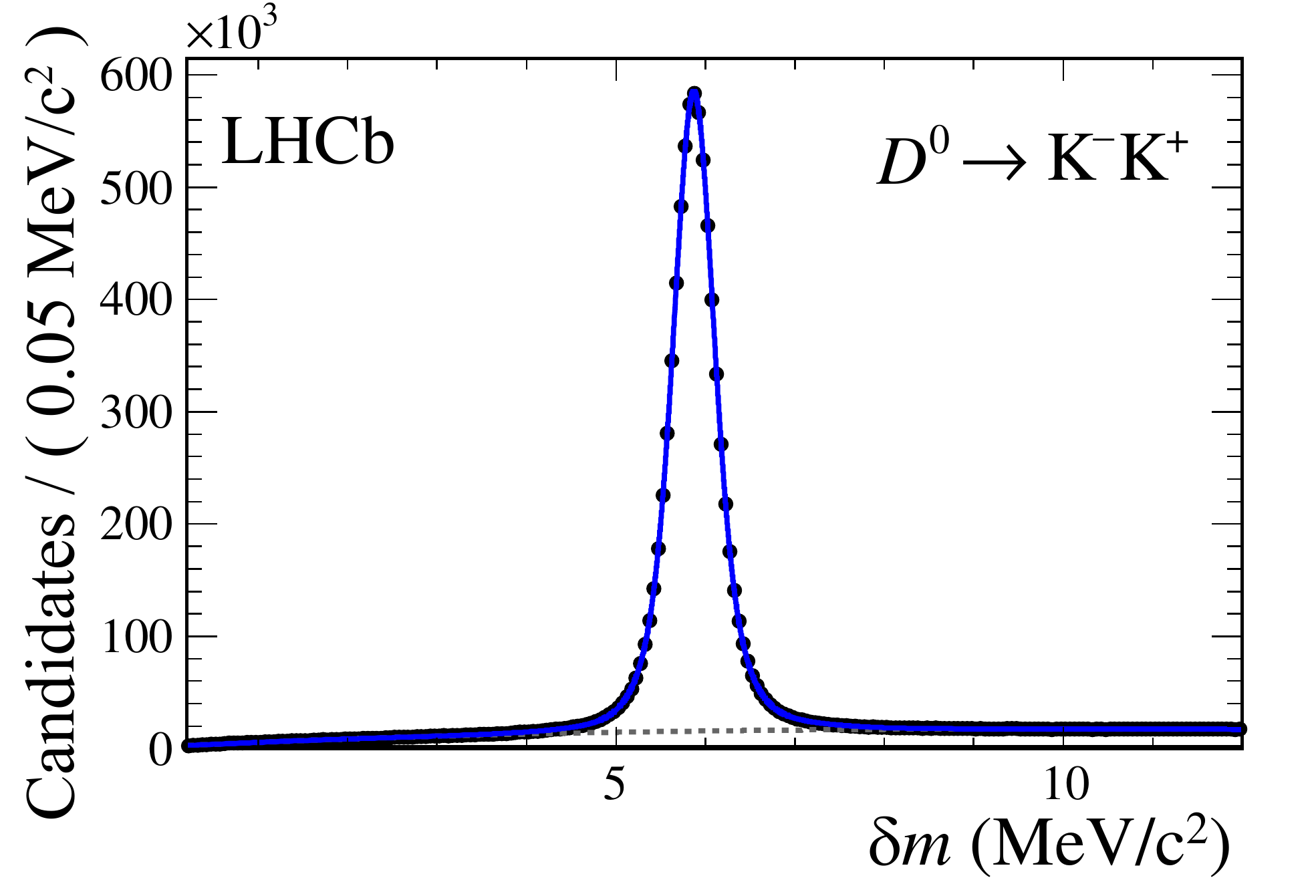}
\includegraphics[width=0.48\textwidth]{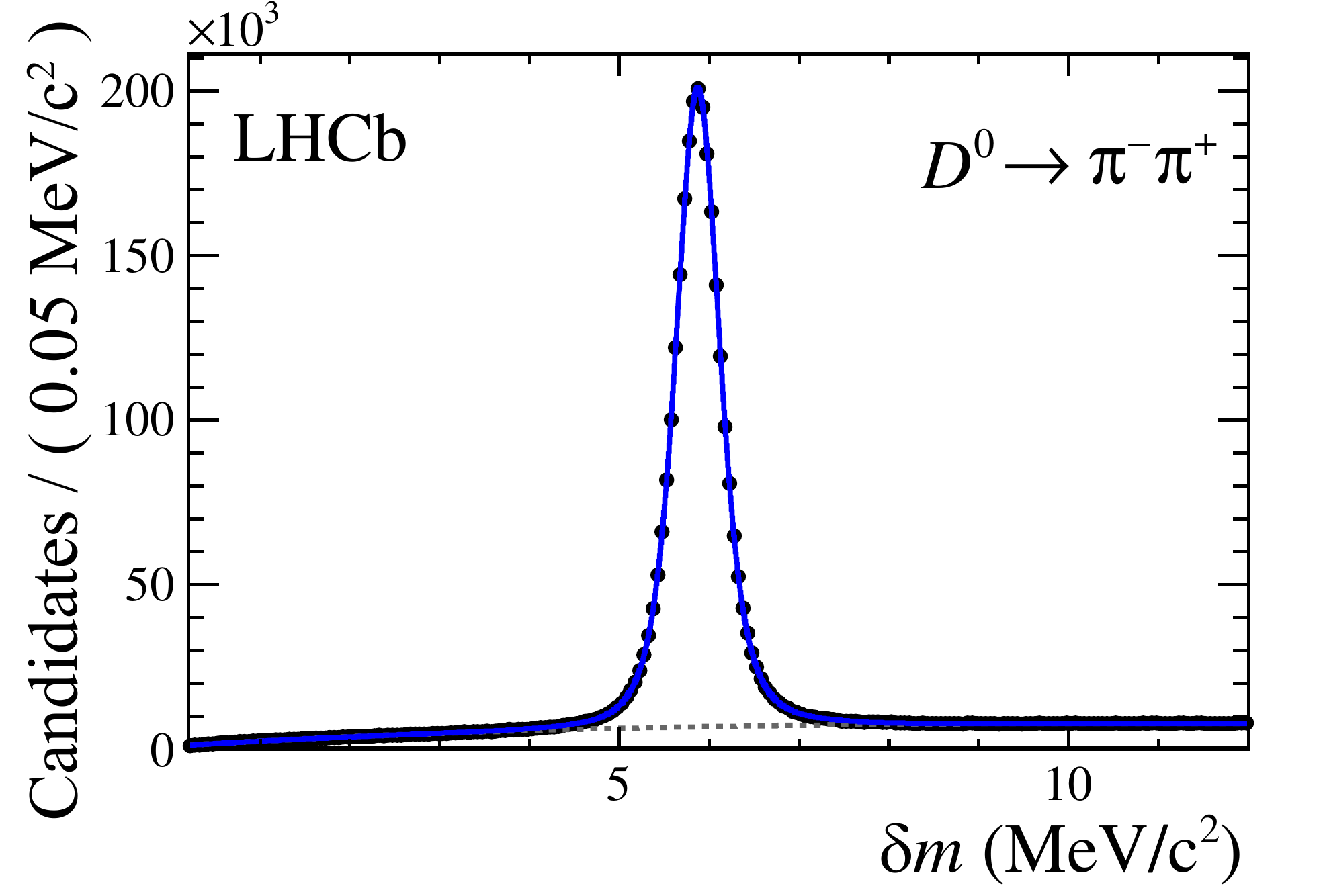}
}

\end{center}
\caption{Fit to the \deltam spectra, where the \Dz is reconstructed in the final state (left) $\Km\Kp$ and (right) $\pim\pip$.
The dashed line corresponds to the background component in the fit.}
\label{fig:dm}
\end{figure}

Signal yields and $\ARAW(\Km\Kp)$ and $\ARAW(\pim\pip)$ are obtained from minimum $\chi^2$ fits to the binned 
$\delta m$ distributions of the \KK and \PiPi samples.
The data samples are split into eight mutually exclusive subsamples separated by centre-of-mass energy, magnet polarity and trigger category.
The signal shape is studied using simulated data and described by the sum of two Gaussian functions with a common mean, and a Johnson $S_{{U}}$ 
function~\cite{Johnson:1949zj}.  The background is described by an empirical function of the form $1-\exp{[(\deltam -\deltam_0)/\alpha]} + \beta (\deltam/\deltam_0 -1)$,
where $\deltam_0$ controls the threshold of the function, and $\alpha$ and $\beta$ describe its shape.
The fits to the eight subsamples and between the \Km\Kp and \pim\pip final states are independent. 
Fits to the $\delta m$ distributions corresponding to the whole data sample are shown in Fig.~\ref{fig:dm}.

The $D^{*+}$ signal yield is $7.7 \times 10^6$ for $\D^0 \rightarrow K^-K^+$ decays,
and $2.5 \times 10^6$ for $\D^0 \rightarrow \pi^-\pi^+$ decays. The signal purity is  $(88.7\pm0.1)\%$ for \KK candidates, and $(87.9\pm0.1)\%$ for \PiPi candidates, in a range of \deltam corresponding to 4.0--7.5\mevcc.
The fits do not distinguish between the signal and the backgrounds that peak in \deltam.  Such backgrounds, which can arise from \Dstarp decays where the correct soft pion is found but the \Dz meson is misreconstructed, are suppressed by the PID requirements to less than 4\% of the number of signal events in the case of $\D^0 \rightarrow K^-K^+$ decays and to a negligible level in the case of \PiPi decays. Examples of such backgrounds are $\Dstarp \to \Dz(\Km\pip\piz)\,\pi^{+}_{s}$ and $\Dstarp \to \Dz(\pim e^{+} \nu_{e} )\,\pi^{+}_s$ decays. The effect on \DACP of residual peaking backgrounds is evaluated as a systematic uncertainty. 

The value of \DACP is determined in each subsample (see Table~\ref{tab:results_all} in Ref.~\cite{supplemental}). Testing the eight independent measurements for mutual consistency gives $\chisq/\rm{ndf} = 6.2/7$, corresponding to a $p$-value of 0.52.
The weighted average of the values corresponding to all subsamples is calculated as $\DACP = (-0.10 \pm 0.08) \%$, where the uncertainty is statistical.

The central value is considerably closer to zero than $\DACP =( -0.82\pm0.21 )\%$, obtained in our previous analysis where a data sample corresponding to an integrated luminosity of 0.6\invfb was considered~\cite{LHCb-PAPER-2011-023}.
Several factors contribute to the change, including the increased size of the data sample  and changes in the detector calibration and reconstruction software. 
To estimate the impact of processing data using different reconstruction software, the data used in Ref.~\cite{LHCb-PAPER-2011-023} are divided into three samples.
The first (second) sample contains events that are selected when using the old (new) version of the reconstruction software and are discarded by the new (old) one, while the third sample consists of those events that are selected by both the versions. 
The measured values are $\DACP = (-1.10\pm0.46) \%$, $\DACP = (0.13\pm0.37)\%$ and $\DACP = (-0.71\pm0.26)\%$, respectively. 
The measurement obtained using the additional data based on an integrated luminosity of 2.4\invfb corresponds to a value of $\DACP = (-0.06 \pm 0.09)\%$. A comparison of the four independent measurements gives $\chisq/\rm{ndf} = 10.5/3$, equivalent to a $p$-value of 0.015. Although this value is small, no evidence of incompatibility among the various sub-samples has been found. Only statistical uncertainties are considered in this study.   

Many sources of systematic uncertainty that may affect the determination of \DACP are considered.  
The possibility of an incorrect description of the signal mass model is investigated by 
replacing the function in the baseline fit with alternative models that provide equally good descriptions of the data.
A value of 0.016\% is assigned as systematic uncertainty, corresponding to the largest variation observed using the alternative functions.   
  
To evaluate the systematic uncertainty related to the presence of multiple candidates in an event, \DACP is measured in samples where one candidate per event is randomly selected. This procedure is repeated one hundred times with a different random selection. The difference of the mean value of these measurements from the nominal result, 0.015\%, is taken as systematic uncertainty.

A systematic uncertainty associated with the presence of background peaking in the $\delta m$ signal distribution and not in the \Dz invariant mass distribution is determined by measuring \DACP from fits to the \Dz invariant mass spectra instead of \deltam. Fits are made for $\Dz \to \Km\Kp$ and $\Dz \to \pim\pip$ candidates within a \deltam window 4.0--7.5\mevcc. 
The background due to genuine \Dz mesons paired with unrelated pions originating from the interaction vertex is subtracted by means of analogous fits to the candidates in the \deltam window 8.0--12.0\mevcc, where the signal is not present. The difference in the \DACP value from the baseline, 0.011\%, is assigned as a systematic uncertainty. 
A systematic uncertainty of 0.004\% is assigned  
for uncertainties associated with the weights calculated for the kinematic reweighting procedure.  
   
A systematic uncertainty is associated with the choice of fiducial requirements on the soft pion applied to exclude regions with large raw asymmetries. To evaluate this uncertainty, the baseline results are compared to results obtained when looser fiducial requirements are applied. The resulting samples include events closer to the regions with large raw asymmetries, at the edges of the detector acceptance and around the beam pipe (see Fig.~\ref{fig:fidcuts1} in Ref.~\cite{supplemental}). The difference in the \DACP values, 0.017\%, is taken as the systematic uncertainty.
 
Although suppressed by the requirement that the \Dz trajectory points back to the primary vertex, \Dstarp mesons produced in the decays of beauty hadrons (secondary charm decays) are still present in the final sample. 
As the $\Dz \to \Km\Kp$ and $\Dz \to \pim\pip$ decays may have different amounts of this contamination, the value of \DACP may be biased because of an incomplete cancellation of the production asymmetries of beauty and charm hadrons. The fractions of secondary charm decays are estimated by performing a fit to the distribution of IP $\chisq$ of the \Dz with respect to all PVs in the event, and are found to be (2.8 $\pm$ 0.1)\% and (3.4 $\pm$ 0.1)\% for the $\Dz \to \Km\Kp$ and $\Dz \to \pim\pip$ samples, respectively. 
Using the LHCb measurements of production asymmetries~\cite{LHCb-PAPER-2014-053, LHCb-PAPER-2014-042, LHCb-PAPER-2012-026, LHCb-PAPER-2015-032}, the corresponding systematic uncertainty is estimated to be 0.004\%.

To investigate other sources of systematic uncertainty, numerous robustness checks have been made. The value of $\Delta A_{\CP}$ is studied as a function of data taking periods and no evidence of any dependence is found. A measurement of \DACP using more restrictive PID requirements is performed, and all variations of \DACP are found to be compatible within statistical uncertainties. To check for possible reconstruction biases, the stability of \DACP is also investigated as a function of many reconstructed quantities, including: the number of reconstructed PVs; the \Dz invariant mass; the \Dz transverse momentum; the \Dz flight distance; the \Dz azimuthal angle; the smallest IP $\chisq$ impact parameter of the \Dz and of the soft pion with respect to all the PVs in the events; the quality of \Dstarp vertex; the transverse momentum of the soft pion; and the quantity $\Delta R = \sqrt{\Delta \phi^2+\Delta \eta^2 }$, where $\Delta \phi$ and $\Delta \eta$ are the differences between \Dz and soft pion azimuthal angles and pseudorapidities. No evidence of dependence of \DACP on any of these variables is found. 
An additional cross check concerns the measured value of $\Delta A_{\rm bkg}$, defined as the difference between the background raw asymmetries $A_{\rm bkg} (\Km\Kp)$ and $A_{\rm bkg} (\pim\pip)$. A value of $\Delta A_{\rm bkg} = (-0.46 \pm 0.13) \%$ is obtained from the fits. In the absence of misidentified or misreconstructed backgrounds, one would expect a value consistent with zero. Decays of \KK and \PiPi have different sources of backgrounds that do not peak in \deltam. These include three-body decays of charmed hadrons with misidentified particles in the final state, as well as four-body decays where one particle is not reconstructed. More restrictive PID requirements have been applied to suppress such backgrounds, and the region of the fits has been extended up to 16\mevcc to improve the precision.
 A value of $\Delta A_{\rm bkg} = (-0.22 \pm 0.13) \%$ is found. The corresponding \DACP value is  $(-0.12 \pm 0.09) \%$, consistent with the baseline result when the overlap of the two samples is taken into account.
Hence, the measurement of \DACP is robust and is not influenced by the background asymmetry. All contributions are summed in quadrature to give a total systematic uncertainty of $0.03\%$.

\begin{figure}[!htb]
  \begin{center}
\includegraphics[width=0.7\textwidth]{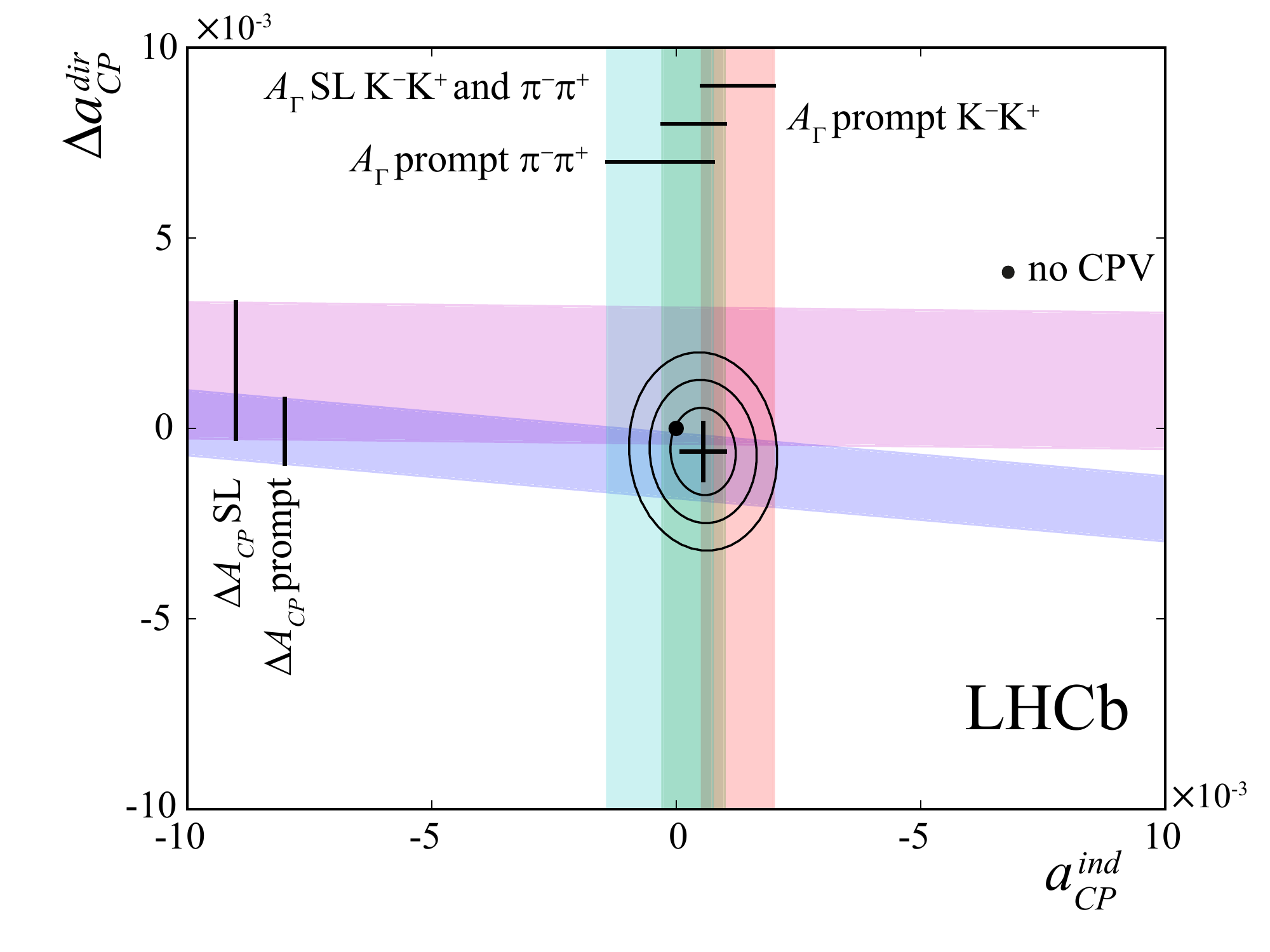}
\end{center}
  \caption{Contour plot of $\Delta a_{\CP}^{\rm dir}$ versus $a_{\CP}^{\rm ind}$. The point at (0,0) denotes the hypothesis of no \CP violation. The solid bands represent the measurements in Refs.~\cite{LHCb-PAPER-2014-013, LHCb-PAPER-2013-054, LHCb-PAPER-2014-069} and the one reported in this Letter. The value of $y_\CP$ is taken from Ref.~\cite{LHCb-PAPER-2011-032}. The contour lines shows the 68\%, 95\% and 99\% confidence-level intervals from the combination.  
}  
  \label{fig:lhcbaverages}
\end{figure}

To interpret the \DACP result in terms of direct and indirect \CP violation, the reconstructed decay time averages, for \KK and \PiPi  samples, are measured. The difference and the average of the mean decay times 
relative to the \Dz lifetime are computed, giving $\Delta\mean{t}/\tau(\Dz)  =  0.1153 \pm 0.0007 \stat \pm 0.0018 \syst $ and  $\overline{\mean{t}}/\tau(\Dz)  =  2.0949 \pm 0.0004 \stat \pm 0.0159 \syst$. The systematic uncertainties are due to the uncertainty on the world average of the \Dz lifetime~\cite{Agashe:2014kda}, decay-time resolution model, and the presence of secondary $\Dz$ mesons from \bquark-hadron decays. 
Given the dependence of \DACP on the direct and indirect \CP asymmetries (Eq.~\ref{eq:dacpdef}) and the measured value  of $\Delta \langle t \rangle / \tau$, the contribution from indirect \CP violation is suppressed and \DACP is primarily sensitive to direct \CP violation. 
Assuming that indirect \CP violation is independent of the \Dz final state, and combining the measurement reported in this Letter with those reported in Ref.~\cite{LHCb-PAPER-2014-013} and with the LHCb measurements of indirect \CP asymmetries ($A_\Gamma \simeq - a_{\CP}^{\rm ind} $)~\cite{LHCb-PAPER-2013-054, LHCb-PAPER-2014-069} and $y_\CP$~\cite{LHCb-PAPER-2011-032}, the values of the direct and indirect \CP asymmetries are found to be 
$a_{\CP}^{\rm ind} = (0.058 \pm 0.044)\%$ and  $\Delta a_{\CP}^{\rm dir} = (-0.061 \pm 0.076)\%$.
Results are summarized in the ($\Delta a_{\CP}^{\rm dir}$, $a_{\CP}^{\rm ind}$) plane shown in  Fig.~\ref{fig:lhcbaverages}. The result is consistent with the hypothesis of \CP symmetry with a $p$-value of 0.32.

In summary, the difference of time-integrated \CP asymmetries between \Dz\to\Km\Kp and \Dz\to\pim\pip decays is measured using $pp$ collision data corresponding to an integrated luminosity of $3.0 \invfb$. The final result is 
\begin{equation}
\DACP = \left(-0.10 \pm 0.08\,\stat  \pm 0.03\,\syst \right)\%, \nonumber
\end{equation}
which supersedes the previous result obtained using the same decay channels based on an integrated luminosity of 0.6\invfb~\cite{LHCb-PAPER-2011-023}. This is the most precise measurement
of a time-integrated \CP asymmetry in the charm sector from a single experiment.

\section*{Acknowledgements}
%The text below are the acknowledgements as approved by the collaboration
%board. Extending the acknowledgements to include individuals from outside the
%collaboration who have contributed to the analysis should be approved by the
%EB. The extra acknowledgements are normally placed before the standard 
%acknowledgements, unless it matches better with the text of the standard 
%acknowledgements to put them elsewhere. They should be included in the draft 
%for the first circulation. Except in exceptional circumstances, to be approved by the
%EB chair, authors of the paper should not be named in extended acknowledgements.
\noindent We express our gratitude to our colleagues in the CERN
accelerator departments for the excellent performance of the LHC. We
thank the technical and administrative staff at the LHCb
institutes. We acknowledge support from CERN and from the national
agencies: CAPES, CNPq, FAPERJ and FINEP (Brazil); NSFC (China);
CNRS/IN2P3 (France); BMBF, DFG and MPG (Germany); INFN (Italy); 
FOM and NWO (The Netherlands); MNiSW and NCN (Poland); MEN/IFA (Romania); 
MinES and FANO (Russia); MinECo (Spain); SNSF and SER (Switzerland); 
NASU (Ukraine); STFC (United Kingdom); NSF (USA).
We acknowledge the computing resources that are provided by CERN, IN2P3 (France), KIT and DESY (Germany), INFN (Italy), SURF (The Netherlands), PIC (Spain), GridPP (United Kingdom), RRCKI and Yandex LLC (Russia), CSCS (Switzerland), IFIN-HH (Romania), CBPF (Brazil), PL-GRID (Poland) and OSC (USA). We are indebted to the communities behind the multiple open 
source software packages on which we depend.
Individual groups or members have received support from AvH Foundation (Germany),
EPLANET, Marie Sk\l{}odowska-Curie Actions and ERC (European Union), 
Conseil G\'{e}n\'{e}ral de Haute-Savoie, Labex ENIGMASS and OCEVU, 
R\'{e}gion Auvergne (France), RFBR and Yandex LLC (Russia), GVA, XuntaGal and GENCAT (Spain), The Royal Society, Royal Commission for the Exhibition of 1851 and the Leverhulme Trust (United Kingdom).

\addcontentsline{toc}{section}{References}
\setboolean{inbibliography}{true}
\bibliographystyle{LHCb}
\bibliography{main,LHCb-PAPER,LHCb-CONF,LHCb-DP,LHCb-TDR}
\clearpage
{\noindent\bf\Large Supplemental material}
\appendix

\begin{figure}[!htb]
  \begin{center}
\includegraphics[width=0.49\textwidth]{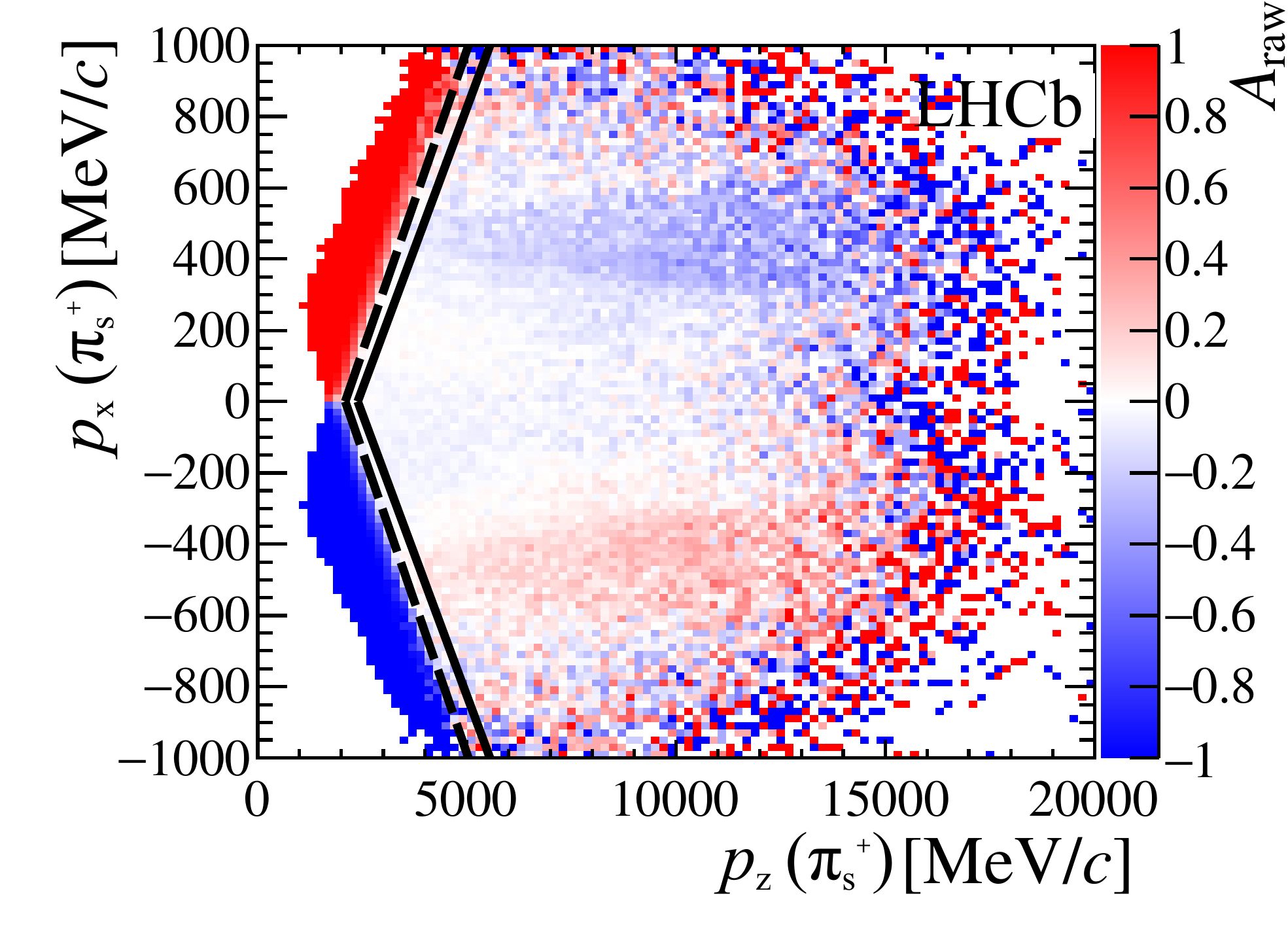}
\includegraphics[width=0.49\textwidth]{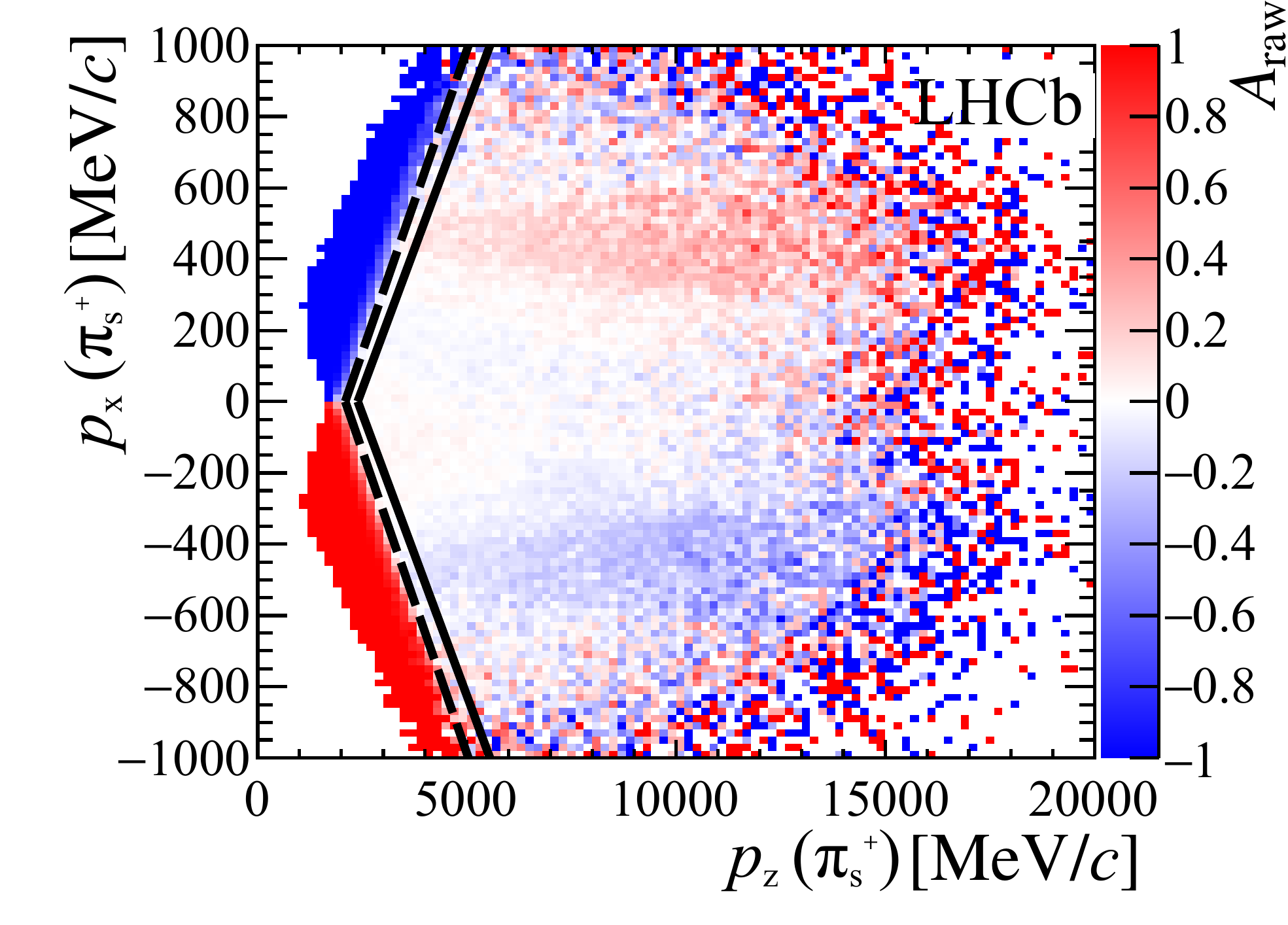}
\end{center}
  \caption{The raw asymmetry, \ARAW, in bins of ($p_z$, $p_x$) of the soft pion with the polarity of the magnet up (left) and down (right). The solid lines show the boundaries corresponding to the baseline selection. The dashed lines represent the edges of the selection used for systematic uncertainty estimation. Candidate \KK decays are shown as an example; however a similar distribution is obtained for \PiPi decays\protect\footnotemark.}
  \label{fig:fidcuts1}
\end{figure}

\footnotetext{The LHCb coordinate system is a right-handed coordinate system, with the $z$ axis pointing along the beam axis, $y$ the vertical direction, and $x$ the horizontal direction. The $(x,z)$ plane is the bending plane of the dipole magnet.}

\begin{figure}[!htb]
  \begin{center}
\includegraphics[width=0.49\textwidth]{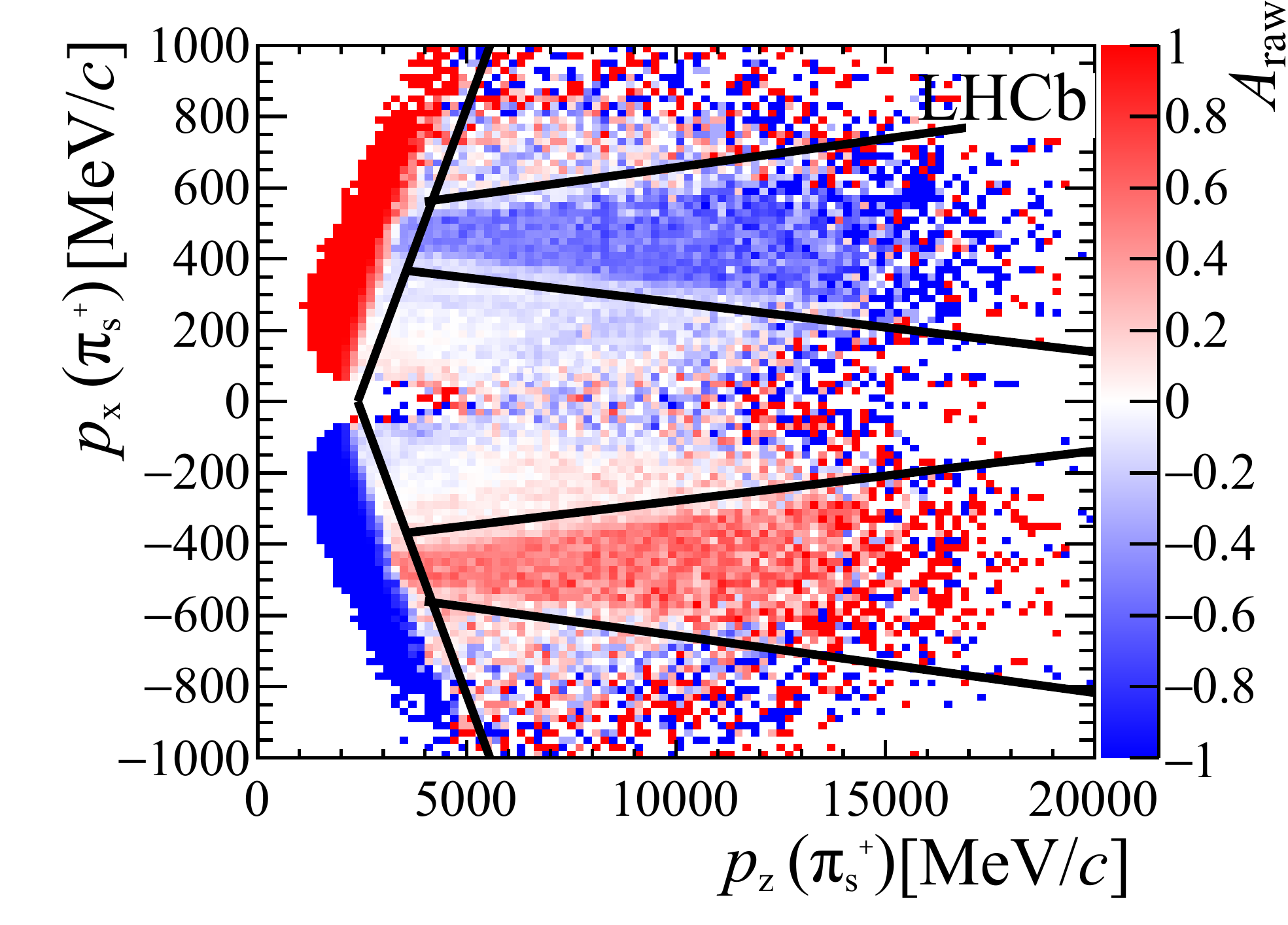}
\includegraphics[width=0.49\textwidth]{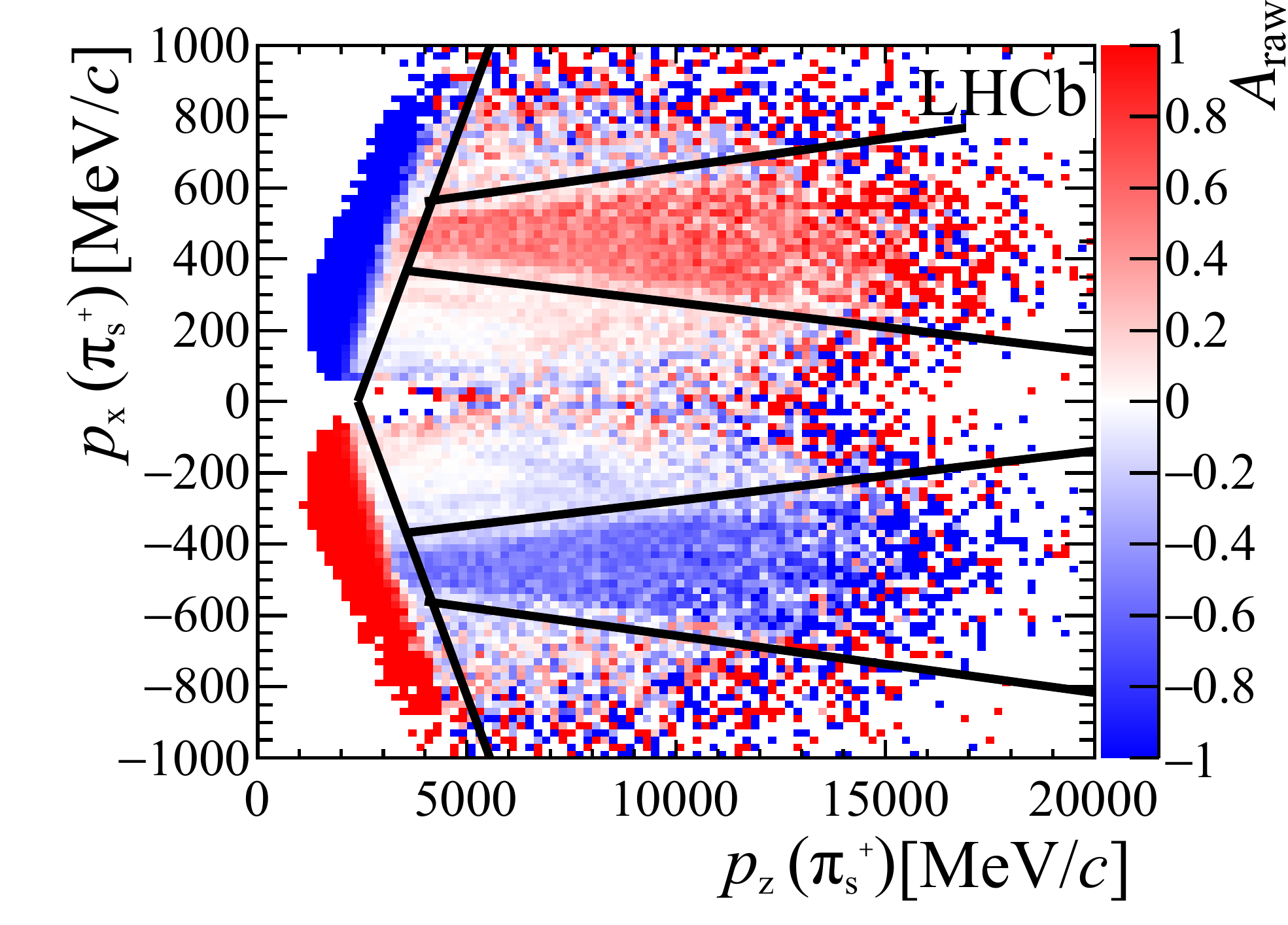}
\end{center}
  \caption{The raw asymmetry, \ARAW, in bins of ($p_z$, $p_x$) of the soft pion with the up (left) and down (right) polarity of the magnetic field. The fiducial requirements (non-instrumented beam pipe region) are superimposed as solid black lines. Only candidates close to the beam pipe region are shown. Candidate \KK decays are shown as an example; however a similar distribution is obtained for \PiPi decays.}
  \label{fig:fidcuts2}
\end{figure}

\begin{table}[tb]
\caption{Values of \DACP measured in the disjoint data subsamples, according to magnet polarity (up, down), centre-of-mass energy of data taking ($\sqrt{s} = 7$ and $8$ TeV ) and trigger category (nTOS, TOS).} 
\begin{center}
\ifthenelse{\boolean{wordcount}}{}{
\begin{tabular}{llcc}
                polarity& trigger & $\sqrt{s}$ [\tev] &\DACP [\%]  \\ \hline
up& TOS& $7$ &$-0.40 \pm  0.35$ \\
up& nTOS& $7$ & $-0.19 \pm  0.29$ \\
down& TOS& $7$ & $-0.31 \pm  0.29$ \\
down& nTOS& $7$ & $-0.06 \pm  0.24$ \\
up& TOS& $8$ & $-0.11 \pm  0.21$ \\
up& nTOS& $8$ & $-0.22 \pm  0.17$ \\
down& TOS& $8$ & $-0.22 \pm  0.21$ \\
down& nTOS& $8$ & $+0.24 \pm  0.17$ \\
\hline
average  & && $-0.10 \pm 0.08$ \\
\end{tabular}
}
\end{center}
\label{tab:results_all}
\end{table} 

\clearpage

\newpage

%%%%%%%%%%%%%%%%%%%%%%%%%%%%%%%%%%%%%%%%%%
\centerline{\large\bf LHCb collaboration}
\begin{flushleft}
\small
R.~Aaij$^{39}$, 
C.~Abell\'{a}n~Beteta$^{41}$, 
B.~Adeva$^{38}$, 
M.~Adinolfi$^{47}$, 
A.~Affolder$^{53}$, 
Z.~Ajaltouni$^{5}$, 
S.~Akar$^{6}$, 
J.~Albrecht$^{10}$, 
F.~Alessio$^{39}$, 
M.~Alexander$^{52}$
% ...

{\footnotesize \it
% ...
$ ^{5}$Clermont Universit\'{e}, Universit\'{e} Blaise Pascal, CNRS/IN2P3, LPC, Clermont-Ferrand, France\\
$ ^{6}$CPPM, Aix-Marseille Universit\'{e}, CNRS/IN2P3, Marseille, France\\
% ...
$ ^{10}$Fakult\"{a}t Physik, Technische Universit\"{a}t Dortmund, Dortmund, Germany\\
% ...
$ ^{38}$Universidad de Santiago de Compostela, Santiago de Compostela, Spain\\
$ ^{39}$European Organization for Nuclear Research (CERN), Geneva, Switzerland\\
% ...
$ ^{41}$Physik-Institut, Universit\"{a}t Z\"{u}rich, Z\"{u}rich, Switzerland\\
% ...
$ ^{47}$H.H. Wills Physics Laboratory, University of Bristol, Bristol, United Kingdom\\
% ...
$ ^{52}$School of Physics and Astronomy, University of Glasgow, Glasgow, United Kingdom\\
$ ^{53}$Oliver Lodge Laboratory, University of Liverpool, Liverpool, United Kingdom\\
% ...
}
\end{flushleft}
%%%%%%%%%%%%%%%%%%%%%%%%%%%%%%%%%%%%%%%%%%

\end{document}